\documentclass[conference]{IEEEtran}
\usepackage{amsmath}
\usepackage{color}
\usepackage{amsthm} 
\usepackage{amssymb}
\usepackage{float}
\usepackage{nccmath}
\usepackage{graphicx}
\usepackage{kotex}
\usepackage[linesnumbered,ruled]{algorithm2e}
\usepackage{setspace}
\usepackage{booktabs}
\usepackage{subcaption}
\usepackage{mwe}
\usepackage{cite}
\usepackage{amsmath,amssymb,amsfonts}
\usepackage{graphicx}
\usepackage{textcomp}
\usepackage{xcolor}

\usepackage{algorithmicx}

\usepackage{xspace}

\begin{document}
\title{Economic Theoretic LEO Satellite Coverage Control: An Auction-based Framework}

\author{\IEEEauthorblockN{$^{\circ}$Junghyun Kim, $^{\S}$Thong D. Ngo, $^{\S}$Paul S. Oh, $^{\S}$Sean S.-C. Kwon, $^{\dag}$Changhee Han, and $^{\circ,\ddag}$Joongheon Kim}
\IEEEauthorblockA{$^{\circ}$School of Electrical Engineering, Korea University, Seoul, Republic of Korea 
\\$^{\S}$Department of Electrical Engineering, California State University, Long Beach, CA, USA
\\$^{\dag}$Artificial Intelligence R\&D Center, Korea Military Academy, Seoul, Republic of Korea \\$^{\ddag}$Artificial Intelligence Engineering Research Center, College of Engineering, Korea University, Seoul, Republic of Korea
\\
E-mails: \texttt{sbdak0312@korea.ac.kr}, 
\texttt{thong.ngo@student.csulb.edu},
\texttt{paul.oh@student.csulb.edu},\\
\texttt{sean.kwon@csulb.edu},
\texttt{chhan46@gmail.com},
\texttt{joongheon@korea.ac.kr}
}
}
\maketitle

\begin{abstract}

Recently, ultra-dense low earth orbit (LEO) satellite constellation over high-frequency bands has considered as one of promising solutions to supply coverage all over the world. 
Given satellite constellations, efficient beam coverage schemes should be employed at satellites to provide seamless services and full-view coverage. 
In LEO systems, hybrid wide and spot beam coverage schemes are generally used, where the LEO provides a wide beam for large area coverage and additional several steering spot beams for high speed data access. 
In this given setting, scheduling multiple spot beams is essentially required. In order to achieve this goal, Vickery-Clarke-Groves (VCG) auction-based trustful algorithm is proposed in this paper for scheduling multiple spot beams for more efficient seamless services and full-view coverage.

\end{abstract}
\IEEEpeerreviewmaketitle
\section{Introduction}
Even though communication and networking technologies are rapidly developing, there is still a long way to go to connect any users at any times in any places.
According to the fact that satellite communications have abundant radio frequency resources, large coverage areas, long communication distance, fast deployment, and less interference from the ground networks are available~\cite{1}. 
The satellite communications have attracted a lot of attention to provide global coverage and satisfy the deficiency of internet access. Due to the distance between the satellite and the surface of the earth, the satellite communications can be classified into geostationary earth orbit (GEO), medium earth orbit (MEO), and low earth orbit (LEO) satellite systems~\cite{1,2}. 
Different form GEO and MEO, LEO satellites are usually located at orbits with altitude from $500$ to $2,000$\,km. Because of the relatively lower latency and path loss, the production and launching costs of LEO are very attractive. To supply worldwide coverage via LEO satellites, we propose a novel hybrid wide and spot beam coverage control scheme is proposed in this paper~\cite{1,3}.

The main idea of the hybrid wide-and-spot beam coverage control is that each LEO satellite provides a wide beam for the total service area as well as several steering spot beams for specific scheduled user terminals. 
The spot beams are always steered to the users, so its trajectory is fixed during the movement of satellite. The power allocation of a spot beam is designed to be much higher than that of the wide beam because the spot beam has to work for modulation/coding techniques to send data to its scheduled user terminals. While a hybrid wide and spot beam coverage control scheme is used, there is a resource allocation problem among user terminals and spot beams~\cite{1},
In general, the number of user terminals demanding data resource allocation is more than the number of available spot beams. 
In this system, a network management center (NMC) which performs resource allocation strategies of the entire network by collecting the traffic data of users and LEO satellites has to effectively allocate spot beams to user terminals. In this paper, a resource assignment method has been proposed that is inspired by the auction which guarantees optimal total utility of the system as well as truthful operations~\cite{4,ref1,ref2,ref3}. The Vickrey-Clarke-Groves (VCG) auction, one of the well-known auction algorithms, is used in this paper because it supposed to be a sealed-bid auction that ensures trustful auctions and guarantees trustful spot beam allocation.

The rest of this paper is organized as follows. 
Sec.~\ref{sec:sec2} presents the details of the proposed auction-based hybrid wide-and-spot beam scheduling and coverage control algorithm. 
Sec.~\ref{sec:sec3} evaluates the performance of the proposed algorithm.
Lastly, Sec.~\ref{sec:sec4} concludes this paper  and presents future research directions.

\section{Auction-based Hybrid Wide/Spot Beam Scheduling and Coverage Control}\label{sec:sec2}

The proposed auction-based hybrid beam scheduling algorithm works based on following three-step procedure. 

\subsection{(Step 1) Auction Setting and Formulation} 
The deployed user terminals in the network system ultimately send their gathered data demand to NMC. NMC calculates the resource allocation among the spot beams and user terminals. As LEO orbits regularly, it is possible to predict spot beams that can be available during given time periods. Note that the future available spot beams are named to FASBs in this paper. Then, FASBs become auctioneers (also called seller) in VCG auction. When the user terminal (the bidders) collect the total data demand rate of the surrounding network, we subtract the total data demand rate from the spot beam's maximum data capacity to measure the spare amount of each spot beam selection data capacity (SBSDC). We can measure the effectiveness of allocation by SBSDC. When SBSDC (i.e., bid) is low, the allocation is regarded as well designed. NMC allocates spot beams to user terminals through an auction mechanism. Although the same user terminals bid to get radio resources from FASBs, user terminals bid differently to each spot beams due to different time, each FASB becomes available. 

In order to formulate the auction mechanism for winning bid determination can based on the definitions as follows.
First of all, the $M$ number of user terminals (bidder) is denoted by
\begin{equation}
\mathcal{U}\triangleq \left\{u^{1},u^{2},\cdots,u^{M}\right\},
\end{equation}
and $N$ number of FASBs (auctioneer) is also denoted by
\begin{equation}
\mathcal{S}\triangleq \left\{s^{1},s^{2},\cdots,s^{N}\right\},
\end{equation}
and lastly, the spare amount of each spot beam data capacity (i.e., bid) is similarly denoted by 
\begin{equation}
\mathcal{B}\triangleq \left\{b_{M}^{1},b_{M}^{2},\cdots,b_{M}^{N}\right\}.
\end{equation}

Then, our mathematical optimization can be formulate as follows. 

\begin{equation}
\min: \sum_{i=1}^{M} \sum_{j=1}^{N} x_{i}^{j}b_{i}^{j} 
\end{equation}
subject to
\begin{eqnarray}
\sum_{j=1}^{N} x_{i}^{j} &\leq& 1, \forall u^{i,j}\in\mathcal{U} \label{eq:p1-const1}\\
\sum_{i=1}^{M} x_{i}^{j} &\leq& 1, \forall s^{i,j}\in\mathcal{S} \label{eq:p1-const2}
\end{eqnarray}

\subsection{(Step 2) Winning Bid Determination}
For winning bid determination, the proposed scheme uses the Hungarian method~\cite{4}. 
The Hungarian method is one of well-known methods for finding a one-to-one matching that meets the minimum bid in the $n \times n$ matrix.
Since the bid matrix is not a square matrix, the NMC needs to modify the bid matrix to used this Hungarian method.
NMC can create a bid matrix of $M \times N$ for each time. 
In this case, if $M > N$, it cannot directly use the Hungarian method. 
The pre-processing for making $n \times n$ matrix can be completed via the following steps.

\begin{enumerate}
\item NMC $g_{j}$ creates a bid matrix of $M \times N$ with $\mathcal{B}$.
\item If $M > N$, add the columns from $L+1$ to $N$ to construct an $M \times N$ matrix.
\item Set any positive integer $Z$ greater than any bids for an element of (1: $M$, $N$+1: $M$).
\item After generating a matrix of $M \times M$, find the minimum total bid and matching through the Hungarian method.
\item Subtract $Z \times (M - (N + 1))$ from the total bid and excluding the device matched with the $N+1$ to $M$-th column.
\end{enumerate}

\subsection{(Step 3) Payment Determination} 
According to the basic theory of VCG auction, the payment for winning bid is determined by the concept of opportunity cost~\cite{ref2,ref3}. The opportunity cost in this system can be determined by $\mathcal{C}_\mathcal{B}$ and $\mathcal{C}_{{\mathcal{B}}\backslash\left\{b_{i}^{j}\right\}}$ where $\mathcal{C}_{\mathcal{B}}$ stands for the summation of determined winning bids by NMC. In addition, $\mathcal{C}_{\mathcal{B}\backslash\left\{b_{i}^{j}\right\}}$ stands for the minimum total cost which is determined by \textit{winning bid determination (in Step 2)} except $b_{i}^{j}$. 
For example, if $b_{i}^{j}$ is a winning bid, it is true that $\mathcal{C}_{\mathcal{B}\backslash\left\{b_{i}^{j}\right\}}> \mathcal{C}_{\mathcal{B}}$. On the other hand, i.e., if $b_{i}^{j}$ is not a winning bid,  $\mathcal{C}_{\mathcal{B}} = \mathcal{C}_{\mathcal{B}\backslash\left\{b_{i}^{j}\right\}}$.
Note that $\mathcal{C}_{\mathcal{B}}-x_{i}^{j}b_{i}^{j}$ is the total cost (determined by winning bid) except its own bid. The payment (i.e., opportunity cost) $p_{i}^{j}$ can be defined as the difference between $\mathcal{C}_{\mathcal{B}\backslash\left\{b_{i,j}^{k}\right\}}$ and  
$\mathcal{C}_{\mathcal{B}}-x_{i}^{j}b_{i}^{j}$~\cite{6}; and thus its mathematical model should be expressed as follows:
\begin{equation}
p_{i}^{j}=
\mathcal{C}_{\mathcal{B}\backslash\left\{b_{i}^{j}\right\}}
-
\left(\mathcal{C}_{\mathcal{B}}-x_{i}^{j}b_{i}^{j}\right).
\label{eq:eq13}
\end{equation}

\section{Performance Evaluation}\label{sec:sec3}

\begin{figure}[htbp]
\begin{center}
    \includegraphics[scale=0.6
    ]{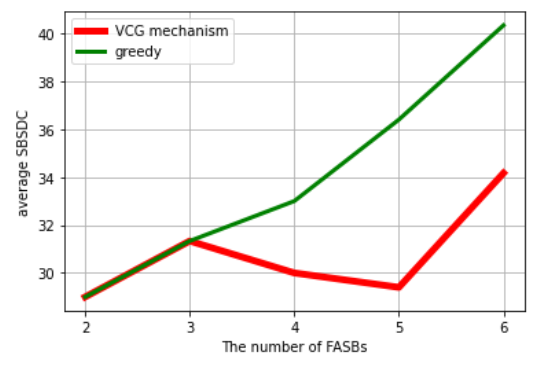}
    \caption{comparative graph of VCG and greedy mechanism} \label{fig:fig1}
\end{center}
\end{figure}

In Fig.~\ref{fig:fig1}, $x$-axis and $y$-axis denotes the number of FASBs that must be assigned to the user terminals and the average SBSDC for spot beams, respectively. 
In general, one spot beam provides data at hundreds of megabits/s. 
As presented in Fig.~\ref{fig:fig1}, the average of 150 megabits/s is assumed as the maximum transmission capacity of each spot beam. VCG auction mechanism uses SBSDCs as bids to find a combination that seeks a minimum total value as mentioned earlier and assigns spot beams to user terminals. Comparatively, the greedy method is to allocate the spot beam by selecting the user terminal that presents the smallest SBSDC whenever a new assignable FASB occurs. The lower values in SBSDC (i.e., $y$-axis value), the more efficient scheduling between the spot beam and the user terminal can be realized. Therefore, we can confirm that VCG auction is more efficient than the greedy method. The reason why the assignable spot beams of two and three produce the same graph values in both policies is that the small number of assignable spot beams results in the same allocation results in the Hungarian and greedy policies. 

\section{Conclusions and Future Work}\label{sec:sec4}
This paper proposes a novel hybrid wide-and-spot beam scheduling for LEO satellite coverage control algorithm inspired by VCG auction mechanism. 
By using the VCG auction, truthful and optimized scheduling for hybrid wide-and-spot beam scheduling can be realized in LEO satellite networks. 
As presented in performance evaluation results, our proposed algorithm achieves desired performance. 

Our potential future work directions include (i) more realistic satellite environment consideration, (ii) more specific system-wide operation definition, and (iii) more specific and data-intensive performance evaluation. 

\section*{Acknowledgment}
This research was supported by the MSIT (Ministry of Science and ICT), Korea, under the ITRC (Information Technology Research Center) support program (IITP-2020-2017-0-01637) supervised by the IITP (Institute for Information \& Communications Technology Planning \& Evaluation). All authors have equal contributions (first authors). S. Kwon, C. Han, and J. Kim are the corresponding authors of this paper.


\begin{thebibliography}{1}

\bibitem{1}
Y. Su, Y. Liu, Y. Zhou, J. Yuan, H. Cao, and J. Shi, ``Broadband LEO Satellite Communications: Architectures and Key Technologies," \textit{IEEE Wireless Communications}, vol. 26, no. 2, pp. 55--61, April 2019.

\bibitem{2}
B. Di, H. Zhang, L. Song, Y. Li, and G.Y. Li, , ``Ultra-Dense LEO: Integrating Terrestrial-Satellite Networks Into 5G and Beyond for Data Offloading," \textit{IEEE Transactions on Wireless Communications}, vol. 18, no. 1, pp. 47--62, January 2019.

\bibitem{3}
Z. Yang, Y. Li, P. Yuan, and Q. Zhang, ``TCSC: A Novel File Distribution Strategy in Integrated LEO Satellite-Terrestrial Networks," \textit{IEEE Transactions on Vehicular Technology}, vol. 69, no. 5, pp. 5426--5441, May 2020.

\bibitem{4}
J. Huang, Z. Han, M. Chiang, and H. V. Poor, ``Auction-Based Resource Allocation for Cooperative Communications," \textit{IEEE Journal on Selected Areas in Communications}, vol. 26, no. 7, pp. 1226--1237, September 2008.

\bibitem{ref1}
M. Shin, J. Kim, and M. Levorato, ``Auction-Based Charging Scheduling With Deep Learning Framework for Multi-Drone Networks," \textit{IEEE Transactions on Vehicular Technology}, vol. 68, no. 5, pp. 4235--4248, May 2019.

\bibitem{ref2}
S. Jeong, W. Na, J. Kim, and S. Cho, ``Internet of Things for Smart Manufacturing System: Trust Issues in Resource Allocation," \textit{IEEE Internet of Things Journal}, vol. 5, no. 6, pp. 4418--4427, December 2018.

\bibitem{ref3}
L. Park, S. Jeong, J. Kim, and S. Cho, ``Joint Geometric Unsupervised Learning and Truthful Auction for Local Energy Market," \textit{IEEE Transactions on Industrial Electronics}, vol. 66, no. 2, pp. 1499--1508, February 2019.


\bibitem{6}
X. Wang, X. Chen, and W. Wu, ``Towards Truthful Auction Mechanisms for Task Assignment in Mobile Device Clouds," in \textit{Proc. IEEE International Conference on Computer Communications (INFOCOM)}, Atlanta, GA, USA, 1-4 May 2017.




\end{thebibliography}
\end{document}